\documentclass[prl,twocolumn,aps,floats,epsf,epsfig,nofootinbib]{revtex4}

\usepackage{graphicx}% Include figure files
\usepackage{dcolumn}% Align table columns on decimal point
\usepackage{bm}% bold math

\def\be{\begin{eqnarray}}
\def\en{\end{eqnarray}}

\def\CP{{\it CP}~}
\def\cp{{\it CP}}

\begin{document}

%\renewcommand{\baselinestretch}{1.10}
%\font\el=cmbx10 scaled \magstep2{\obeylines\hfill May, 2011}

%\vskip 1.5 cm

\title{ Wolfenstein Parametrization at Higher Order: \\
Seeming Discrepancies and Their Resolution }

\author{ Y. H. Ahn, Hai-Yang Cheng, and Sechul Oh }

\address{Institute of Physics, Academia Sinica  \\
Taipei, Taiwan 115, Republic of China}

%\medskip

\date{\today}

\begin{abstract}
The precise determination of the Cabibbo-Kobayashi-Maskawa (CKM) matrix elements is extremely important towards
the understanding of {\it CP} violation.
We explicitly study seeming discrepancies between the CKM matrix elements at the higher order of the
expansion parameter $\lambda$ in {\it different} Wolfenstein parametrizations derived from different
exact parametrizations.  A systematic way of resolving the
seeming discrepancies is proposed.  We find that most of the discrepancies can be naturally resolved by
a proper redefinition of the numerically small (of order $\lambda$) parameters.
Our approach is further applied to the cases for the Wolfenstein-{\it like} parametrizations, such
as the Qin-Ma parametrization.
\end{abstract}

\maketitle

%%%%%%%%%%%%%%%%%%%%%%%%%%%%%%%%%%%%%%%%%%%%%
%
%  Text
%
%%%%%%%%%%%%%%%%%%%%%%%%%%%%%%%%%%%%%%%%%%%%%

{\it CP} violation plays a key role in explaining the observed matter-antimatter asymmetry in our universe which is one of the greatest unsolved puzzles in physics. Hence, a full understanding of the underlying mechanism for {\it CP} nonconservation is one of the hottest issues in
modern physics. In the Standard Model (SM), the origin of {\it CP} violation resides solely in the phase in the Cabibbo-Kobayashi-Maskawa
(CKM) matrix which is a $3 \times 3$ one indicating three generations of quarks in Nature.
The CKM matrix elements are the fundamental parameters in the SM, the {\it precise} determination of which is
highly crucial and will be performed in future experiments such as LHCb and Super $B$ factory ones.
Apparently, if the CKM matrix is
expressed in a particular parametrization, such as the Wolfenstein one, having an approximated form
in terms of a small expansion parameter $\lambda$, then high order $\lambda$ terms in the CKM
matrix elements to be determined in the future precision experiments will become more and more important.

Physics should be independent of a particular parametrization of the CKM matrix.
Owing to its practical usefulness and importance, the Wolfenstein parametrization of the CKM matrix has
been one of the most popular parametrizations since its first appearance in 1983~\cite{Wolfenstein:1983yz}.
It was pointed out~\cite{Wolfenstein_Buras} that as in any perturbative expansion, high order terms
in $\lambda$ are not unique in the  Wolfenstein parametrization, though the nonuniqueness of the high
order terms does not change the physics.
Thus, if one keeps using only one parametrization, there would not be any problem.
However, if one tries to compare the values of certain parameters, such as $\lambda$, used in
one parametrization with those used in another parametrization, certain complications can occur
(as we shall see later), because of the nonuniqueness of the high order terms in $\lambda$.
Since the CKM matrix can be parametrized in infinitely many ways with three rotation angles and one
\cp-odd phase, it is desirable to find a certain systematic way to resolve these complications and to
keep consistency between the CKM matrix elements expressed in different parametrizations.
In this work, we explicitly explore the seeming discrepancies between the CKM matrix elements at the
high order of $\lambda$ in {\it different} Wolfenstein parametrizations obtained from different exact
parametrizations.  Then we propose a systematic way of resolving the seeming discrepancies.
In particular, we shall see that most of the discrepancies can be naturally resolved by a proper
redefinition of the numerically small parameters.
Our approach is then extended to the cases for the Wolfenstein-{\it like} parametrizations.

The magnitudes of the CKM matrix elements determined by global fits are~\cite{PDG10}
\be   \label{eq:Vfit}
&& \hspace{-0.5cm} V =
  \left( \matrix{
  |V_{ud}| & |V_{us}| & |V_{ub}|  \cr
  |V_{cd}| & |V_{cs}| & |V_{cb}|  \cr
  |V_{td}| & |V_{ts}| & |V_{tb}|  \cr}\right)  \\
&& \hspace{-0.5cm} =
\footnotesize
  \left( \matrix{
  0.97428\pm 0.00015 & 0.2253\pm{0.0007} &0.00347^{+0.00016}_{-0.00012} \cr \cr
  0.2252\pm0.0007  & 0.97345^{+0.00015}_{-0.00016} & 0.0410^{+0.0011}_{-0.0007} \cr \cr
  0.00862^{+0.00026}_{-0.00020}  & 0.0403^{+0.0011}_{-0.0007} &
  0.999152^{+0.000030}_{-0.000045} \cr}\right).  \nonumber
\en
Among many possibilities of parametrizations of $V$, the well-known Kobayashi-Maskawa (KM) [or CKM]
parametrization is given by~\cite{Kobayashi:1973fv}
\be \label{eq:KM}
\footnotesize
V_{\rm KM} = \left( \matrix{
  c_1,  &  -s_1 c_3,  &  -s_1 s_3  \cr \cr
  s_1 c_2,  &  c_1 c_2 c_3  &  c_1 c_2 s_3  \cr
    & ~ -s_2 s_3 e^{i\delta_{\rm KM}},  &  ~ +s_2 c_3 e^{i\delta_{\rm KM}}  \cr \cr
  s_1 s_2,  &  c_1 s_2 c_3  &  c_1 s_2 s_3  \cr
    & ~ +c_2 s_3 e^{i\delta_{\rm KM}},  &  ~ -c_2 c_3 e^{i\delta_{\rm KM}} }\right),
\en
where $c_i \equiv \cos\theta_i$ and $s_i \equiv \sin\theta_i$.
On the other hand, the Chau-Keung (CK) [or Chau-Keung-Maiani ($\overline{\rm CKM}$)]
parametrization~\cite{Chau:1984fp} has been advocated by the Particle Data Group
(PDG)~\cite{PDG10} to be the standard parametrization for the quark mixing matrix:
\be  \label{eq:CK}
\footnotesize
V_{\rm CK}
 = \left( \matrix{
  c_{12} c_{13},  &  s_{12} c_{13},  &  s_{13} e^{-i\phi}  \cr \cr
  -s_{12} c_{23}  &  c_{12} c_{23}  &  s_{23} c_{13}  \cr
    -c_{12} s_{23} s_{13} e^{i\phi},  &  -s_{12} s_{23} s_{13} e^{i\phi},  &   \cr \cr
  s_{12} s_{23}  &  -c_{12} s_{23}  &  c_{23} c_{13} \cr
    -c_{12} c_{23} s_{13} e^{i\phi},  &  -s_{12} c_{23} s_{13} e^{i\phi},  &   \cr }
 \right).
\en

Before proceeding it should be stressed that there exist nine fundamentally different ways for describing the CKM  matrix \cite{Fritzsch:1997st,Rasin:1997pn}). (Of course, the freedom of rotating the phase of quark fields will render the parametrization
of the quark mixing matrix infinitely many.) Among them, the most popular ones are the KM, CK and Fritzsch-Xing (FX) \cite{Fritzsch:1997fw} parametrizations. Although these different parametrizations are mathematically equivalent, they have a different theoretical motivation and one of them may turn out to be more convenient for some specific problem. For example, the imaginary part appears in the CK parametrization with a smaller coefficient of order $10^{-3}$, contrary to the KM one where the imaginary part of the matrix element, e.g. $V_{tb}$, is large and comparable to the real part. The FX parametrization is motivated
by the hierarchical structure of the quark masses. It has primarily a heavy
quark mixing involving the $t$ and $b$ quarks whereas the {\it CP}-odd phase resides solely in the light quark sector \cite{Fritzsch:1997fw}. On the other hand, it is known that among the possible parametrizations of the CKM matrix, only the KM and FX ones can allow to have maximal \CP violation \cite{Koide:2004gj}, namely, the phase $\delta_{\rm KM}$ in the KM parametrization (see Eq. (\ref{eq:KM})) is in the vicinity of $90^\circ$.

The magnitudes of the CKM matrix given in Eq. (\ref{eq:Vfit}) show a hierarchical pattern
with the diagonal elements being close to unity, the elements $|V_{us}|$ and $|V_{cd}|$ being of order 0.23, the elements $|V_{cb}|$ and $|V_{ts}|$ of order 0.04 whereas $|V_{ub}|$ and $|V_{td}|$ are of order $(3-9)\times 10^{-3}$. The Wolfenstein parametrization given in Eq. (\ref{eq:Wolf}) below exhibits this hierarchy manifestly and transparently. Moreover, the imaginary parts are suppressed as they first appear at order $\lambda^3$. The transparency of the Wolfenstein form and its smallness of {\it CP} violation explains why this parametrization is so popular and successful in the phenomenological applications.
It is an approximate parametrization of the CKM matrix expanded as a power series in terms of the small parameter $\lambda\approx |V_{us}|$; the three angles and one phase in various exact parametrizations are replaced by the four real parameters $\lambda,~A,~\rho$ and $\eta$. A new Wolfenstein-{\it like} parametrization has been advocated recently by Qin and Ma (QM) \cite{Qin:2010hn} in which the three angles are substituted by the parameters  $\lambda,~f$ and $h$ while the phase parameter $\delta$ is still kept.  Unlike the original Wolfenstein parametrization, the QM one has the advantage that its {\it CP}-odd phase $\delta$ is manifested in the parametrization and close to $90^\circ$ [see Eq. (\ref{eq:QMfh}) below]. In a recent work, we have shown that this feature of maximal {\it CP} violation is crucial for a viable neutrino phenomenology \cite{Ahn:2011ep}.

The Wolfenstein parametrization~\cite{Wolfenstein:1983yz} was introduced as
\be \label{eq:Wolf}
\footnotesize
V_{\rm Wolf} = \left( \matrix{
  1 -\frac{1}{2} \lambda^2 ,  & \lambda ,  & A \lambda^3 (\rho-i\eta  \cr
    &   &  +i \eta \frac{1}{2} \lambda^2)   \cr \cr
  -\lambda ,  & 1 -\frac{1}{2} \lambda^2   & A \lambda^2 (1 +i \lambda^2 \eta) \cr
    &  -i \eta A^2 \lambda^4 ,  &   \cr \cr
  A \lambda^3 (1 -\rho -i\eta),  & -A \lambda^2 ,  & 1  \cr
 } \right) ,
\en
where it was demanded that the imaginary part of the unitarity relation be satisfied to order $\lambda^5$
and the real part only to order $\lambda^3$.  It was noted in~\cite{Wolfenstein:1983yz} that the term
$i \eta A \lambda^5 /2$ in $V_{ub}$ could be transferred to $V_{td}$.
Using the global fits to the data, the four unknown real parameters $A$, $\lambda$, $\rho$ and $\eta$
are determined to be~\cite{PDG10}
\be
&& A = 0.808^{+0.022}_{-0.015}\,, \quad \lambda=0.2253\pm 0.0007\,, \nonumber \\
&& \bar\rho=0.132^{+0.022}_{-0.014}\,,~ \quad \bar\eta=0.341\pm{0.013}\,,
\en
where $\bar \rho=\rho(1-\lambda^2/2+\cdots)$ and $\bar\eta=\eta(1-\lambda^2/2+\cdots)$.
In principle, the expression of the Wolfenstein parametrization to the high order of $\lambda$ can be
systematically obtained from the exact parametrization of the CKM matrix by expanding it to
the desired order of $\lambda$.
It is well known that the Wolfenstein parametrization can be easily obtained from the standard
CK parametrization in Eq.~(\ref{eq:CK}) by applying the relations
\be \label{eq:CKtoW}
 s_{12} = \lambda, \quad s_{23} = A \lambda^2, \quad s_{13}e^{-i\phi} = A \lambda^3 (\rho-i\eta).
\en
The detailed expression up to order $\lambda^6$ is given by
\begin{widetext}
\be   \label{eq:Wolf_CK_HO}
V_{\rm Wolf}^{{\rm (CK)}} &=&
\footnotesize
 \left( \begin{array}{ccc}
  1 -\frac{\lambda^2}{2} -\frac{\lambda^4}{8}
    & \lambda  ~,
    & A \lambda^3 (\rho -i \eta)
  \\
    -\frac{\lambda^6}{16} [1 +8 A^2 (\rho^2 +\eta^2)] ~,
    &
    &
  \\ \\
  -\lambda +\frac{\lambda^5}{2} A^2 (1 -2\rho -2i \eta) ~,
    & 1 -\frac{\lambda^2}{2} -\frac{\lambda^4}{8} (1 +4 A^2)
    & A \lambda^2
  \\
    & ~ -\frac{\lambda^6}{16} [1 -4 A^2 (1 -4\rho -4i\eta)] ~,
    &
  \\ \\
  A \lambda^3 (1 -\rho -i \eta)
    & -A \lambda^2 +\frac{\lambda^4}{2} A (1 -2\rho -2i \eta)
    & 1 -\frac{\lambda^4}{2} A^2
  \\
  +\frac{\lambda^5}{2} A (\rho +i \eta) ~,
    & +\frac{\lambda^6}{8} A ~,
    & -\frac{\lambda^6}{2} A^2 (\rho^2 +\eta^2)
 \end{array} \right)  +{\cal O}(\lambda^7) ~.
\en
\end{widetext}
Here we note that the next higher order $\lambda$ term in $V_{us}$ appears at order $\lambda^7$
~({\it i.e.}, $-\frac{1}{2} A^2 \lambda^7 (\rho^2 +\eta^2)$).
In comparison with the original Wolfenstein form in Eq.~(\ref{eq:Wolf}), the imaginary term
$i \eta A \lambda^5 /2$ has been transferred from $V_{ub}$ to $V_{td}$.
The imaginary terms $i \eta A \lambda^4$ of $V_{cb}$ and $-i \eta A^2 \lambda^4$ of $V_{cs}$ in the
original Wolfenstein form have been transferred to $-i \eta A \lambda^4$ of $V_{ts}$ and
$-i \eta A^2 \lambda^5$ of $V_{cd}$, respectively, which satisfy the unitarity relations.

It has been pointed out~\cite{QinMa_ACO} that the Wolfenstein parametrization can be also obtained from
the KM parametrization in Eq.~(\ref{eq:KM}) by first rotating the phases of some of the quark fields
$s\to s\,e^{i\pi}$,  $c\to c\,e^{i\pi}$, $b\to b\,e^{i(\theta+\pi)}$,
$t\to t\,e^{-i(\delta_{\rm KM}-\theta)}$,
and then using the relations
\be \label{eq:KMtoW}
&& s_1= \lambda', \qquad s_2e^{-i(\delta_{\rm KM}-\theta)}=A' \lambda'^2 (1 -\rho' -i\eta'),
  \nonumber \\
&& s_3 e^{-i\theta}=A' \lambda'^2 (\rho' -i\eta') ~,
\en
where the primed $\lambda$, $A$, $\rho$, $\eta$ are used to distinguish them from the unprimed ones
in $V_{\rm Wolf}^{\rm (CK)}$, as in general the primed parameters can be different from the unprimed ones.
The result up to order $\lambda'^6$ reads
\begin{widetext}
\be   \label{eq:Wolf_KM_HO}
V_{\rm Wolf}^{{\rm (KM)}} &=&
\footnotesize
 \left( \begin{array}{ccc}
  1 -\frac{\lambda'^2}{2} -\frac{\lambda'^4}{8} -\frac{\lambda'^6}{16} ~,
    & \lambda' -\frac{\lambda'^5}{2} A'^2 (\rho'^2 +\eta'^2) ~,
    & A' \lambda'^3 (\rho' -i \eta')
  \\ \\
  -\lambda'
    & 1 -\frac{\lambda'^2}{2}
    & A' \lambda'^2 -\frac{\lambda'^4}{2} A' (\rho' -i\eta')
  \\
  +\frac{\lambda'^5}{2} A'^2 [(1 -\rho')^2 +\eta'^2)] ~,
    & -\frac{\lambda'^4}{8} [ 1 +4 A'^2 (1 +2i \eta') ]
    & -\frac{\lambda'^6}{8} A' \Big[ \rho' +4 A'^2 (\rho' -\rho'^2 +\eta'^2)
  \\
    & -\frac{\lambda'^6}{16} [ 1 -4 A'^2 (1 -2\rho' +2\rho'^2 +2\eta'^2) ] ~,
    & -i\eta' (1 +4 A'^2 (1 -2\rho')) \Big]
  \\ \\
  A' \lambda'^3 (1 -\rho' -i \eta') ~,
    & ~ -A' \lambda'^2 +\frac{\lambda'^4}{2} A' (1 -\rho' -i \eta')
    & ~ 1 -\frac{\lambda'^4}{2} A'^2 (1 -2i\eta')
  \\
    & +\frac{\lambda'^6}{8} A' \Big[ 1 -\rho' +4 A'^2 (\rho' -\rho'^2 +\eta'^2)
    & ~ +\frac{\lambda'^6}{2} A'^2 [\rho' (1 -\rho') -\eta'^2 -i\eta']
  \\
    & -i\eta' (1 -4 A'^2 (1 -2\rho')) \Big]  ~,
    &
 \end{array} \right)  +{\cal O}(\lambda'^7) ~.  \nonumber \\
\en
\end{widetext}
We see that most of the matrix elements of $V_{\rm Wolf}^{\rm (KM)}$ in Eq.~(\ref{eq:Wolf_KM_HO}) look
quite different from those of $V_{\rm Wolf}^{\rm (CK)}$ in Eq.~(\ref{eq:Wolf_CK_HO}),
even though up to order $\lambda^3$, the corresponding elements of the two matrices are exactly same.
For instance, the matrix element $V_{us}$ now has the term of order $\lambda'^5$,
$-\frac{1}{2} A'^2 \lambda'^5 (\rho'^2 +\eta'^2)$, while $V_{td}$ does not have the term of order
$\lambda^5$, $\frac{\lambda^5}{2} A (\rho +i \eta)$ which appears in the counterpart of
Eq.~(\ref{eq:Wolf_CK_HO}).
Of course, physical observables should be determined independently of a particular parametrization of
the CKM matrix so that the magnitude of each matrix element in $V_{\rm Wolf}^{\rm (KM)}$ should be the
same as the corresponding one in $V_{\rm Wolf}^{\rm (CK)}$, even though they look quite different from
each other.
We shall show later that much of these seeming discrepancies can be naturally resolved by a proper
redefinition of the relevant parameters $\rho^{(\prime)}$ and $\eta^{(\prime)}$.

Now let us discuss in detail why the above discrepancies occur in the two Wolfenstein parametrizations
derived from the exact CK and KM parametrizations, respectively.
In Eqs.~(\ref{eq:KM}) and (\ref{eq:CK}), taking into account the fact that $s_i$ and $s_{ij}$ are small
quantities of order $\lambda^m$ with $m > 0$, and $c_i$ and $c_{ij}$ are of order unity, one finds that
$s_1 \approx s_{12} = O(\lambda)$ from $|V_{us}| = O(\lambda)$, and $s_{23} = O(\lambda^2) \approx s_2$
and/or $s_{23} \approx s_3$ from $|V_{cb}| = O(\lambda^2)$.
But, because $|V_{ub}| =s_1 s_3$ in $V_{\rm Wolf}^{\rm (KM)}$,
$s_3 \approx s_{13} / O(\lambda) \gg s_{13}$ indicating that $c_3$ and $c_{13}$ deviate from each
other at the subleading order: subsequently, for example, the discrepancies occur between the high order
terms in $\lambda$ of $V_{us}$ (or $V_{cb}$) in $V_{\rm Wolf}^{\rm (CK)}$ and $V_{\rm Wolf}^{\rm (KM)}$,
respectively.
In addition, the assignment of the \cp-odd phase to different matrix elements in the CK and KM
parametrizations, respectively, ({\it i.e.,} $\delta_{\rm KM}$ being assigned to $V_{cs}$, $V_{cb}$,
$V_{ts}$, $V_{tb}$, but $\phi$ being assigned to $V_{cd}$, $V_{cs}$, $V_{td}$, $V_{ts}$) leads to
different imaginary terms proportional to $i\eta$ (or $i\eta'$) in the two aforementioned Wolfenstein
parametrizations, respectively.

In comparison with the data $|V_{ub}| < 0.2 |V_{cb}| \simeq A \lambda^3$ which Wolfenstein used for his
original parametrization in Eq.~(\ref{eq:Wolf}), the current data shown in Eq.~(\ref{eq:Vfit}) indicates
$|V_{ub}| \sim \lambda^2 |V_{cb}| \simeq A \lambda^4$.  Thus, we propose to {\it define} the parameters
$\tilde \rho$ and $\tilde \eta$ of {\it order unity} by scaling the numerically small (of order $\lambda$)
parameters $\rho$ and $\eta$ as
\be  \label{rho_eta}
\footnotesize
 \tilde \rho \equiv \rho / \lambda ~, \quad \tilde \eta \equiv \eta / \lambda ~, \quad
\tilde \rho' \equiv \rho' / \lambda' ~,  \quad~ \tilde \eta' \equiv \eta' / \lambda' ~,
\en
where the numerical values of $\tilde \rho$ and $\tilde \eta$ are $0.601^{+0.098}_{-0.062}$ and
$1.553^{+0.054}_{-0.055}$, respectively~\cite{PDG10}.
Then $V_{ub}$ becomes of order $\lambda^4$, instead of the conventional order $\lambda^3$, while
$V_{td}$ still has a leading term of order $\lambda^3$.
Consequently, $V_{\rm Wolf}^{\rm (CK)}$ and $V_{\rm Wolf}^{\rm (KM)}$ become
\begin{widetext}
\be   \label{eq:Wolf_CK_HO2}
V_{{\rm Wolf}}^{({\rm CK})} &=&
\footnotesize
 \left( \begin{array}{ccc}
  1 -\frac{\lambda^2}{2} -\frac{\lambda^4}{8} -\frac{\lambda^6}{16} ~,
    & \lambda  ~,
    & A \lambda^4 (\tilde \rho -i \tilde \eta)
    \nonumber \\ \\
  -\lambda +\frac{\lambda^5}{2} A^2 -A^2 \lambda^6 (\tilde \rho +i \tilde \eta) ~,
    & ~ 1 -\frac{\lambda^2}{2} -\frac{\lambda^4}{8} (1 +4 A^2) -\frac{\lambda^6}{16} (1 -4 A^2) ~,
    & A \lambda^2
    \nonumber \\ \\
  A \lambda^3 -A \lambda^4 (\tilde \rho +i \tilde \eta)
    +\frac{\lambda^6}{2} A (\tilde \rho +i \tilde \eta) ~,
    & -A \lambda^2 +\frac{\lambda^4}{2} A -\lambda^5 A (\tilde \rho +i \tilde \eta) +\frac{\lambda^6}{8} A ~,
    & 1 -\frac{\lambda^4}{2} A^2
 \end{array} \right)  +{\cal O}(\lambda^7) ~, \\ \\
V_{{\rm Wolf}}^{({\rm KM})} &=&
\footnotesize
 \left( \begin{array}{ccc}
  1 -\frac{\lambda'^2}{2} -\frac{\lambda'^4}{8} -\frac{\lambda'^6}{16} ~,
    & \lambda' ~,
    & A' \lambda'^4 (\tilde \rho' -i \tilde \eta')
    \nonumber \\ \\
  -\lambda' +\frac{\lambda'^5}{2} A'^2 -A'^2 \lambda'^6 \tilde \rho' ~,
    & 1 -\frac{\lambda'^2}{2} -\frac{\lambda'^4}{8} (1 +4 A'^2) -i A'^2 \lambda'^5 \tilde \eta'
    & A' \lambda'^2 -\frac{\lambda'^5}{2} A' (\tilde \rho' -i \tilde \eta')
    \nonumber \\
    & ~~~~ -\frac{\lambda'^6}{16} (1 -4 A'^2)  ~,
    &
    \nonumber \\ \\
  A' \lambda'^3 -A' \lambda'^4 (\tilde \rho' +i \tilde \eta') ~,
    & ~ -A' \lambda'^2 +\frac{\lambda'^4}{2} A' -\lambda'^5 A' (\tilde \rho' +i \tilde \eta')
      +\frac{\lambda'^6}{8} A' ~,
    & ~ 1 -\frac{\lambda'^4}{2} A'^2 +i A'^2 \lambda'^5 \tilde \eta'
 \end{array} \right)  +{\cal O}(\lambda'^7) ~. \\ \label{eq:Wolf_KM_HO2}
\en
\end{widetext}
Indeed, after the redefinition in Eq.~(\ref{rho_eta}), the seeming discrepancies between the corresponding
elements of $V_{\rm Wolf}^{\rm (CK)}$ in Eq.~(\ref{eq:Wolf_CK_HO}) and $V_{\rm Wolf}^{\rm (KM)}$ in
Eq.~(\ref{eq:Wolf_KM_HO}) are resolved significantly.
Especially $V_{ud}$, $V_{us}$ and $V_{ts}$ are now matched in form in both $V_{\rm Wolf}^{\rm (CK)}$
and $V_{\rm Wolf}^{\rm (KM)}$.
Although there still remain some discrepancies in the matrix elements $V_{cd}$, $V_{cs}$, $V_{cb}$,
$V_{td}$, and $V_{tb}$, most of them arise from the additional imaginary terms proportional
to $i \tilde \eta$ or $i \tilde \eta'$.

In order to resolve the remaining discrepancies between $V_{{\rm Wolf}}^{\rm (CK)}$ in
Eq.~(\ref{eq:Wolf_CK_HO2}) and $V_{{\rm Wolf}}^{\rm (KM)}$ in Eq.~(\ref{eq:Wolf_KM_HO2}),
we further propose a systematic prescription as follows:

\noindent
(i) From $V_{us}$ in both $V_{\rm Wolf}^{\rm (CK)}$ and $V_{\rm Wolf}^{\rm (KM)}$, we {\it define}
\be  \label{pres_lambda}
V_{us} \equiv \lambda =\lambda' ~.
\en
In fact, at any given order $\lambda^n$ with the integer $n > 6$, one can always make $V_{us}$
real and define $V_{us} \equiv \lambda$ in any parametrization: {\it e.g.},
at order $\lambda^9$, $V_{us} = \lambda -\frac{1}{2} A^2 \lambda^9 (\tilde \rho^2 +\tilde \eta^2)$
in $V_{\rm Wolf}^{\rm (CK)}$, but $V_{us} = \lambda' -\frac{1}{2} A'^2 \lambda'^7
(\tilde \rho'^2 +\tilde \eta'^2)$ in $V_{\rm Wolf}^{\rm (KM)}$.
At this given order $\lambda^9$, one can define the whole $V_{us}$ just as a new real $\lambda''$,
being determined by experimental measurements of $V_{us}$, for both $V_{\rm Wolf}^{\rm (CK)}$
and $V_{\rm Wolf}^{\rm (KM)}$. Subsequently the other matrix elements in $V_{\rm Wolf}^{\rm (CK)}$ and
$V_{\rm Wolf}^{\rm (KM)}$ can be recast in terms of $\lambda''$.

\noindent
(ii) From $V_{cb}$ in both $V_{{\rm Wolf}}^{\rm (CK)}$ and $V_{{\rm Wolf}}^{\rm (KM)}$, since only
the magnitude of $V_{cb}$ is a physical observable, we set
\be  \label{pres_AA}
A = A' \left| 1 -\frac{\lambda'^3}{2} (\tilde \rho' -i \tilde \eta') \right| ~.
\en
%Here we note that both terms on the left-hand and right-hand sides are real.

\noindent
(iii) From $V_{ub}$ in both $V_{{\rm Wolf}}^{\rm (CK)}$ and $V_{{\rm Wolf}}^{\rm (KM)}$, we put
\be  \label{pres_rho_eta}
\tilde \rho -i \tilde \eta =(\tilde \rho' -i \tilde \eta')
  \left| 1 -\frac{\lambda'^3}{2} (\tilde \rho' -i \tilde \eta') \right|^{-1} ~.
\en
It is easy to check that the magnitude ({\it i.e.,} physical observable) of each element of
$V_{{\rm Wolf}}^{\rm (CK)}$  is the same as the corresponding one in  $V_{{\rm Wolf}}^{\rm (KM)}$.
Therefore, the discrepancies between the corresponding elements of $V_{{\rm Wolf}}^{\rm (CK)}$ in
Eq.~(\ref{eq:Wolf_CK_HO}) and $V_{{\rm Wolf}}^{\rm (KM)}$ in Eq.~(\ref{eq:Wolf_KM_HO}) can be resolved
through the redefinition of the numerically small (of order $\lambda$) parameters in Eq.~(\ref{rho_eta})
and the prescription given in Eqs.~(\ref{pres_lambda})$-$(\ref{pres_rho_eta}).

The above redefinition and prescription can be also applied to the Wolfenstein-{\it like}
parametrizations~\cite{Qin:2010hn,Kim_Seo}.
As an example, let us consider the Qin-Ma (QM) parametrization~\cite{Qin:2010hn} which is a new
Wolfenstein-{\it like} parametrization based on the triminimal expansion of the CKM matrix.
It is obtained from the KM parametrization in Eq. (\ref{eq:KM}) by first making the phase rotation
$s\to s\,e^{i\pi}$, $c\to c\,e^{i\pi}$, $b\to b\,e^{i(\pi-\delta_{\rm KM})}$,
and then applying the relations
$s_1= \lambda$,  $\qquad s_2=f\lambda^2$ and  $s_3e^{-i\delta_{\rm KM}}=h\lambda^2e^{-i\delta_{\rm QM}}$.
Up to order $\lambda^6$, we obtain
\begin{widetext}
\be  \label{eq:QM_HO}
V_{{\rm QM}}^{\rm (KM)} &=&
\small
\left( \begin{array}{ccc}
1 -\frac{\lambda^2}{2}  -\frac{\lambda^4}{8}  -{\lambda^6\over 16},
  & \lambda,
  & \tilde h\lambda^4 e^{-i \delta_{\rm QM}}  \nonumber \\ \\
-\lambda +\frac{\lambda^5}{2}f^2 ,
  & ~ 1 -\frac{\lambda^2}{2} -\frac{\lambda^4}{8} (1 +4 f^2) -f \tilde h \lambda^5 e^{i \delta_{\rm QM}}
  & ~~ f \lambda^2 +\tilde h \lambda^3 e^{-i \delta_{\rm QM}} -\frac{\lambda^5}{2} \tilde h e^{-i \delta_{\rm QM}}
  \nonumber \\
  & \qquad~~~ +{\lambda^6\over16}(4f^2-4\tilde h^2-1) \ , &   \nonumber \\ \\
f \lambda^3 ,
  & -f\lambda^2 -\tilde h \lambda^3 e^{i\delta_{\rm QM}} +\frac{\lambda^4}{2} f +{\lambda^6\over 8}f ,
  & 1 -\frac{\lambda^4}{2} f^2 -f \tilde h \lambda^5 e^{-i \delta_{\rm QM}} -{\lambda^6 \over 2}\tilde h^2
\end{array} \right)  +{\cal O}(\lambda^7) ~, \nonumber\\
\en
\end{widetext}
where the parameters $A$, $\rho$ and $\eta$ in the Wolfenstein parametrization are replaced by $f$, $h$
and $\delta$. In Eq. (\ref{eq:QM_HO}) we have {\it defined} the parameter of {\it order unity}
$\tilde h\equiv h/\lambda$.
From the global fits to the CKM matrix given in Eq. (\ref{eq:Vfit}), the parameters $f$, $\tilde h$ and
$\delta$ are determined to be
\be \label{eq:QMfh}
&& f = 0.754^{+0.016}_{-0.011}\,, \qquad \tilde h=1.347^{+0.045}_{-0.030}\,, \nonumber \\
&& \delta_{\rm QM}=(90.4^{+0.36}_{-1.15})^\circ \,.
\en
We have shown in \cite{QinMa_ACO} that the QM phase $\delta_{\rm QM}$ is the same as the KM phase
$\delta_{\rm KM}$; they are both approximately maximal.
It is straightforward to show that the matrix elements of $V_{{\rm QM}}^{\rm (KM)}$ are identical in
magnitude to the corresponding ones of $V_{{\rm Wolf}}^{\rm (CK)}$, provided that QM parameters are related
to the Wolfenstein ones through the following relations. We find
(i) the same $\lambda$ from $V_{us}$,
(ii) $\tilde h^2 = A^2 (\tilde\rho^2+\tilde\eta^2)$ from $V_{ub}$,
(iii)
{\footnotesize
\be \label{eq:f2}
f^2 = A^2 \Big[ 1 -2 \lambda \tilde\rho \left( 1 -\frac{\lambda^2}{2} \right)
  +\lambda^2 (\tilde\rho^2 +\tilde\eta^2) \left( 1 -\frac{\lambda^2}{2} \right)^2 \Big]
\en
}
from $V_{td}$, (Since $f$ is of order unity, higher order $\lambda$ terms in Eq. (\ref{eq:f2}) can be
neglected so that $f$ is expanded to order $\lambda^3$.)
and (iv) $A^2=f^2+\tilde h^2\lambda^2(1-\lambda^2/2)^2$ from $V_{cb}$ together with
$\delta_{\rm QM}=90^\circ$. It is easily seen that the relation in (iv) follows from (ii) and (iii)
as a good approximation. These relations are in agreement with those obtained in~\cite{QinMa_ACO} except
for some higher order $\lambda$ corrections.

In conclusion, as the high precision era of the CKM matrix elements comes, we have shown that the
seeming discrepancies between the CKM matrix elements at high order
of $\lambda$ occur in {\it different} Wolfenstein(-{\it like}) parametrizations derived from the exact
CK and KM parametrizations, respectively.
Our systematic prescription can resolve the seeming discrepancies.
Especially, it turns out that most of the discrepancies can be naturally resolved through the definition of
the parameters $\tilde \rho$, $\tilde \eta$, $\tilde h$ of {\it order unity} by scaling the numerically
small (of order $\lambda$) parameters $\rho$, $\eta$, $h$ as
$\tilde \rho \equiv \rho /\lambda$, $\tilde \eta \equiv \eta /\lambda$, $\tilde h \equiv h /\lambda$.

\vskip 0.2cm
{\bf Acknowledgments}

This research was supported in part by the National Science Council of R.O.C. under Grant Nos.
NSC-97-2112-M-008-002-MY3, NSC-97-2112-M-001-004-MY3 and NSC-99-2811-M-001-038.

\end{document}